\documentclass[conference]{IEEEtran}

\usepackage{cite}
\usepackage{amsmath,amssymb,amsfonts}
\usepackage{graphicx}
\usepackage[T1]{fontenc}
\usepackage{textcomp}
\usepackage{xcolor}
\usepackage{subcaption}
\usepackage{hyperref}
\usepackage{algorithm}
\usepackage[noend]{algpseudocode}
\def\BibTeX{{\rm B\kern-.05em{\sc i\kern-.025em b}\kern-.08em
    T\kern-.1667em\lower.7ex\hbox{E}\kern-.125emX}}

\usepackage{graphicx}
\usepackage{subcaption}
\usepackage{mathtools}

\usepackage{booktabs}

\begin{document}
\title{Modeling Pointing, Acquisition, and Tracking Delays in Free-Space Optical Satellite Networks}

\author{
\IEEEauthorblockN{
Jason Gerard\IEEEauthorrefmark{1},
Juan A. Fraire\IEEEauthorrefmark{3},
Sandra Céspedes\IEEEauthorrefmark{1}
}

\IEEEauthorblockA{
\IEEEauthorrefmark{1}Department of Computer Science \& Software Engineering, Concordia University, Montreal, Canada.\\ 
\IEEEauthorrefmark{2}Inria, INSA Lyon, CITI, UR3720, 69621 Villeurbanne, France.\\ 
\IEEEauthorrefmark{3}CONICET - Universidad Nacional de Córdoba, Córdoba, Argentina.\\
}
}


\maketitle

\begin{abstract}
Free-space optical communications enable high-capacity optical inter-satellite links (OISL) and direct-to-Earth (DTE) links, but require precise Pointing, Acquisition, and Tracking (PAT) between terminals.
Current scheduling approaches often overlook or oversimplify PAT delays, leading to inefficient contact planning and overestimated network capacities.
We present a validated model for quantifying PAT delays before data transmission begins, encompassing coarse pointing, fine pointing, and the handover to tracking.
The model is grounded in mission data from NASA TBIRD, LLCD, DSOC, and ESA’s Lunar Optical Communication Terminal.
We find that PAT delays exhibit multimodal distributions based on prior link geometry and scale nonlinearly with initial pointing uncertainty and optical beam width.
Integrating these delay models into routing and scheduling algorithms will enable more accurate contact planning and higher utilization in optical networks.
The proposed model provides a foundation for evaluating performance and designing algorithms for future large-scale optical satellite networks.
\end{abstract}


\section{Introduction}

The space industry is experiencing unprecedented growth in the "New Space" era, with breakthroughs in reusable rockets and CubeSat-standard satellites dramatically lowering exploration costs.
This revolution propels ambitious missions, including NASA's Artemis and Mars Exploration Programs, SpaceX's Starship missions, and ESA's Moonlight project \cite{israel_lunanet_2020}.
These missions transmit large amounts of scientific data through interplanetary network (IPN) links back to Earth.
However, NASA's Deep Space Network (DSN) faces critical limitations, including constrained bandwidth, limited radio frequency bands, and insufficient network autonomy, all of which necessitate manual scheduling intervention.
The capacity shortfall is severe—at certain times, up to 75\% of Mars mission data cannot be transmitted to Earth due to bandwidth constraints~\cite{gladden_lessons_2024}.

Free-space optical (FSO) communications offer a compelling solution to address network capacity limitations by providing significantly higher data rates, ranging from Gbps to Tbps, while requiring lower transmission power due to narrow beam spread \cite{bhattacharjee_ondemand_2024}.
This enables deployment with reduced size, weight, power, and cost (SWaP-C) compared to RF systems.

Although optical communications enable high-data-rate communications, the technology also poses significant challenges due to its reliance on precise pointing, acquisition, and tracking (PAT)~\cite{bhattacharjee_ondemand_2024}.
Unlike IPN links, where PAT delays dominate due to large pointing uncertainties and weaker beacon signals, low Earth orbit (LEO) networks, such as Starlink or Amazon Kuiper, maintain stable mesh topologies with persistent along-track optical inter-satellite links (OISL).
However, cross-plane links and LEO-to-ground contacts still require frequent retargeting and experience non-negligible PAT delays.
Whether for deep space or LEO scenarios, the narrow beam width of the laser demands highly accurate mechanical alignment between the transmitter and receiver, which must be reestablished for each new contact.
This makes link setup sensitive to disruptions in line-of-sight (LOS) conditions and imposes strict requirements on synchronized contact schedules.
Currently, these schedules are often managed manually by mission operations centers (MOCs), adding operational complexity.
Furthermore, existing scheduling and routing algorithms for FSO networks fail to account for the time required by the PAT process, resulting in inefficient contact selection and an overestimation of communication resources~\cite{gerard_autonomous_2024, bhattacharjee_ondemand_2024}.
There is a critical need for models that accurately quantify retargeting frequency, PAT delays, and the resulting transmission duty cycle to better understand and optimize network capacity in optical space networks.


Experimental data from missions such as NASA's TBIRD, LLCD, and LCRD, as well as ESA's Lunar Optical Communication Terminal tests, have demonstrated these delays in practice across LEO, lunar, and deep space environments.
Despite these mission results, there has been limited structured modeling of PAT delays under varying acquisition methods and pointing strategies within the literature.
Developing accurate models of these delays is crucial for algorithmic research in optical network routing and link scheduling, ensuring that PAT delays are properly accounted for during the selection and sequencing of links and paths.
Without robust theoretical models, it is challenging to analyze network throughput, contact planning, and the dynamic behavior of FSO mesh networks and the future solar system internet (SSI) under realistic operational constraints.

\subsection{Related Work}

Prior work often overlooks the dynamic delays introduced by pointing and acquisition.
The authors in~\cite{liang_latency_2023} compute optical link latency as the sum of propagation delays along network paths.
However, in non-stable optical mesh networks, PAT delays can significantly increase total latency.
While initial FSO mesh networks may maintain persistent optical paths, deep space missions with disruptions from orbital occlusions and LEO mega-constellations with multiple shells and optical ground stations will require frequent retargeting, requiring that link setup delays are explicitly incorporated into link scheduling and routing.
Projections of less than two-second link setup delays in future terminals~\cite{chaudhry_temporary_2022} are optimistic, as current deep space missions observe acquisition times closer to 200 seconds~\cite{biswas_deep_2024}.
Ignoring these delays limits the accuracy of network evaluations and highlights the need to explicitly incorporate PAT delays into optical link scheduling and routing.

The authors in~\cite{fraire_optimal_2024} study how effective contact duration, defined as the total contact time minus acquisition delay, is influenced by the placement of the optical head.
They use a static acquisition delay of 100 seconds in their analysis.
While the study shows that optimal placement can significantly improve effective contact duration, it does not consider how placement may positively or negatively affect the duration of PAT itself.
This remains an area of valuable insight that has not yet been explored, largely due to the lack of sufficient theoretical models.
As a result, authors often rely on estimates from similar real-world missions to approximate these delays.

In terms of optical link routing and scheduling, the algorithms proposed in~\cite{gerard_autonomous_2024, bhattacharjee_ondemand_2024} assume a static PAT delay for edge selection.
This assumption does not capture the dynamics of pointing and link acquisition, which in practice vary based on hardware specifications, internal control system algorithms, and stochastic processes.
The algorithms themselves also influence PAT delays, as repeatedly scheduling the same link avoids additional PAT delay.
These variations impact the cost associated with each edge in maximum weight matchings or minimal weight path selections, potentially leading to different routing and scheduling outcomes.
Existing optical link scheduling and routing algorithms inadequately account for the impact of PAT on effective contact duration, resulting in inefficient contact selection.

\subsection{Contributions of this Work}

This work presents a modeling framework for quantifying delays across the full PAT sequence, including coarse pointing, beam searching, and transitioning to tracking.
The model is validated using NASA and ESA mission data, showing optical link characterization based on multimodal PAT delays and nonlinear scaling with the field of uncertainty (FOU).
These findings support the integration of PAT delays into optical network performance analysis, as well as routing and scheduling algorithms, enabling improved planning and utilization in the SSI, composed of LEO, lunar, and deep space FSO networks.

\section{Coarse Pointing between Optical Terminals}

\begin{figure}[t]
    \centering
    \includegraphics[width=0.9\linewidth,height=\textheight,keepaspectratio]{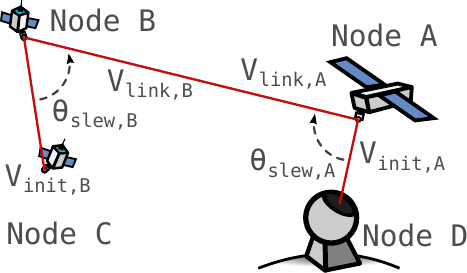}
    \caption{Example of the angle change for Node A when slewing its optical head from an optical ground station to a deep space satellite, Node B.}
    \label{fig:pat_pointing}
\end{figure}

During coarse pointing, the optical head mechanically slews toward the expected position of the partner satellite or optical ground station, as shown in Fig.~\ref{fig:pat_pointing}, to establish the initial LOS for the upcoming contact.
The expected pointing vector is computed using spacecraft cueing data, which includes predicted or real-time ephemeris and attitude information.
Ephemeris provides precise position and velocity data, whereas attitude describes the spacecraft's orientation.

Depending on the terminal architecture, coarse pointing is performed either by rotating the entire spacecraft using reaction wheels, torque rods, and the spacecraft propulsion system, or through a two-axis coarse pointing assembly (CPA).
Each optical head has a field of regard (FOR), which is the maximum angular extent, in the azimuth and elevation directions, through which the CPA can be positioned.
The spacecraft may have to adjust its attitude such that the partner terminal's field of uncertainty (FOU) falls within the FOR.

Pose data, defined as position and attitude estimates in a global reference frame, is critical for computing the LOS vector used to determine the initial pointing direction.
These estimates are derived from GPS receivers, star trackers, gyroscopes, and inertial measurement units, and are combined with propagated orbital ephemeris data.
The onboard software uses the information to compute the expected LOS vector to the partner terminal (or ground station), and transforms it from the inertial frame into the spacecraft body frame to define the desired boresight pointing angle.
Once the estimated boresight arrives at the center of the FOU, the system transitions to the acquisition phase.

The duration of the pointing phase depends on the total angle the optical head must slew to direct the boresight to the FOU, and the slew rate in the azimuth and elevation directions.
Given a 5-minute contact window, as seen in NASA's TBIRD mission~\cite{riesing_operations_2023}, the slew can take up to 15\% of the total duration, making a noticeable impact on the amount of data that can be transferred during the contact window.

\subsection{Pointing Vector Delay Model}

The proposed delay model uses pointing vectors that represent directional edges between optical terminals to compute the duration of the slew required to transition from the previous to the next partner terminal.

For each link, an optical terminal $n$ has two 3-dimensional pointing vectors, representing where the optical head is currently pointing, $V_{init,n}$, and where it will point to for the link, $V_{link,n}$. This is illustrated in Fig.~\ref{fig:pat_pointing} by vectors $V_{init,A}$ and $V_{link,A}$ in Node A. Each pointing vector can be split into its azimuth and elevation components, as the CPA will generally have two different slew rates.
The origin of the pointing vector is the center point of the satellite's optical head at $<0,0,0>$.

Figure~\ref{fig:pat_pointing} shows $\theta_{slew,n}$, the angle between the initial pointing vector of the current link at $t=1$ and the next link at $t=2$.
The slew angle can be decomposed into $\theta_{slew,n,az}$ and $\theta_{slew,n,el}$.
The angle is defined as
\begin{equation}
    \theta_{slew,n} = \arccos(\frac{V_{init,n} \cdot V_{link,n}}{||V_{init,n}|| \bullet ||V_{link,n}||}).
\end{equation}
Therefore, the pointing delay is defined as $T_{\mathit{pointing}} = \frac{\theta_{slew,n}}{\omega_{n}}$, where $\omega_{n}$ is the angular velocity, also called the slew rate, of the CPA of spacecraft $n$ in either the azimuth, $\omega_{az}$, or elevation, $\omega_{el}$, direction.

The target and partner optical terminals can only begin link acquisition when both nodes have finished coarse pointing.
The pointing delay varies between nodes as the number of degrees required to retarget depends on the position of the previous partner satellite.
To account for this restriction, we compute $T_{\mathit{pointing}}$ as the maximum value between the pointing delays of each side of the link.

\subsection{Pointing Error \& Calibration}

Due to uncertainties in attitude estimation, thermal expansion, launch vibration-induced misalignments, and ephemeris errors, the resulting alignment typically brings the optical beam within a broad uncertainty cone, known as the FOU, around the target direction.
This cone may range from milliradians to microradians, depending on the precision of the ephemeris data and pointing error of the CPA.
For example, NASA’s TBIRD mission cited inaccurate TLEs, which are used to derive the initial pointing uncertainty as the primary mode of contact failure~\cite{riesing_operations_2023}.

Mechanical pointing errors are counteracted through a combination of optical head calibration and fine beam alignment using the fast steering mirrors (FSM).
Post-launch calibration aligns the optical head using celestial references to correct initial pointing offsets before first contact.
Pre-link calibration homes the optical head and aligns the FSM for boresight correction periodically after initial commissioning.
Throughout PAT, the system computes the point-ahead angle for the transmit beam to compensate for the one-way light time delay between the transmitter and receiver, which occurs due to the high relative velocities of spacecraft in orbit.
This angle is applied through a separate point-ahead assembly or through a combination of multiple FSMs that preserve the bidirectional capabilities of the optical terminal.

\section{Optical Link Acquisition}

During link acquisition, the optical terminal actively searches for and locks onto the incoming signal from the partner terminal.
This process includes fine pointing, beam searching, and the transition to closed-loop tracking.
Each of these sub-processes is critical to ensuring that the receiver can detect, acquire, and maintain a high optical signal-to-noise ratio (OSNR) link.



\subsection{Fine Pointing \& Beam Searching}

Fine pointing begins once the coarse pointing system has aligned the terminal at the center of the FOU, the expected acquisition region.
A fine-pointing mechanism, such as an FSM, is used to refine the beam position within the narrow field of view (FOV) of the receiver, i.e., the viewable angular area seen through the optical telescope.
FSMs electronically point the optical beam by tipping and tilting the mirror, allowing for fine angular adjustments, typically measured in microradians per second.
Fine pointing enables sub-beamwidth tracking accuracy during the acquisition and closed-loop tracking phases.

The system uses a track sensor that takes feedback from a light detector, typically a quadrant photodiode, to determine the offset between the incoming signal centroid and the center of the optical axis.
This offset is fed into a high-bandwidth control loop, which computes actuator commands to continuously correct for small angular deviations. NASA previously demonstrated closed-loop fine tracking using quad-cell feedback and FSM control during the LLCD mission~\cite{boroson_overview_2014}.

\begin{figure}[t]
    \centering
    \includegraphics[width=1.0\linewidth,height=\textheight,keepaspectratio]{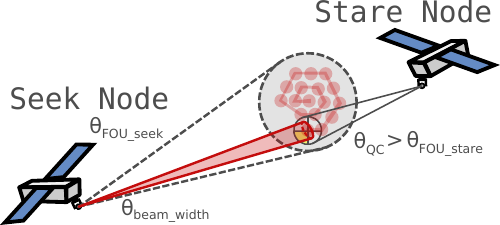}
    \caption{The seek satellite performing a beam search using a hexagonal spiral while the stare satellite detected the angle of incidence of the light using the quad-cell based track sensor.}
    \label{fig:pat_acq_beam_search}
\end{figure}

To reduce the complexity of link acquisition, only a single optical terminal performs the beam search, also known as seeking, while the other optical terminal stares at the FOU.
To this end, the stare node's track sensor FOV, $\theta_{QC}$, should be larger than the FOU, $\theta_{FOU\_stare}$, as illustrated in Fig.~\ref{fig:pat_acq_beam_search}.
This ensures that the seeking node's laser or beacon remains within the track sensor's FOV during the fine pointing process.
The seeking terminal begins the search at the center point of the seeking node's FOU, $\theta_{FOU\_seek}$.
If the incoming beacon is not immediately visible within the detector's FOV, the system initiates a beam search pattern to scan the FOU.
Once the seeking terminal, shown in Fig.~\ref{fig:pat_acq_beam_search}, detects light in its track sensor, it shines light back at the seeking terminal, indicating that the two terminals have locked on to each other.

The search patterns are designed to maximize coverage of the search cone while minimizing acquisition time.
Figure~\ref{fig:pat_acq_beam_search} shows the common spiral search pattern where the beam is moved in an expanding circular pattern through the FOU.
Other patterns are also used, such as raster scans, which step the beam in a grid-like pattern, and hybrid or rosette patterns.
NASA’s DSOC payload on the Psyche mission uses spiral scanning during initial downlink acquisition from deep space, enabling it to find the Earth-based receiver despite large angular uncertainties~\cite{desoria-santacruz_systems_2024}.

Some optical systems further reduce the duration of link acquisition by adjusting the beam divergence---starting with a wide beam to quickly scan the field of uncertainty (FOU), then switching to a narrow beam for efficient data transmission. Optical ground stations often take this approach, as they are not limited by energy constraints and do not incur additional launch costs due to weight.
Deep space links can opt for a single laser for both beaconing and data transmission, requiring additional steps during the beam search process.
Beam searching can be bypassed when using a star tracker–only approach, which depends on precise positioning, navigation, timing (PNT), and pointing hardware. However, this method may not be feasible for all deep space missions~\cite{swank_beaconless_2016}. RF-assisted FOU reduction can also eliminate the need for beam searching~\cite{fernandez-nino_rfassisted_2024}.

Following the seek-stare approach and a hexagonal spiral pattern, the duration of the beam search is only dependent on the seeking terminal.
To compute the duration of the structured search across the FOU, $T_{\mathit{seek}}$, we propose first computing the number of steps, $N_{steps}$, required to fill the FOU based on the beam width: 
\begin{equation} \label{eq:n_steps}
N_{steps} = \left\lceil \frac{2\pi R^2}{\sqrt{3} d^2} \right\rceil,
\end{equation}
where $R$ is the radius of the FOU and $d$ is the beam width (diameter).
Both $R$ and $d$ can be computed using $D \cdot \tan\left(\frac{\theta}{2}\right)$, where $\theta$ is the angle.
In (~\ref{eq:n_steps}), the numerator represents the area of the FOU, while the denominator is based on hexagonal close-packing geometry, which calculates the area covered by each beam spot in a hexagonal grid---the most efficient way to tile circles in two dimensions
$N_{steps}$ is the worst-case number of steps needed to detect and acquire the signal of the partner terminal within the FOU, given that it must search through the entire pattern.


In a hexagonal pattern, we observe four diagonal slews, given by $N_{\mathit{d}} = 4 \cdot \left\lceil \frac{N_{steps}}{6} \right\rceil$ for every two horizontal slews $N_{\mathit{h}} = 2 \cdot \left\lceil \frac{N_{steps}}{6} \right\rceil$.
The total seek time is then defined by
\begin{equation}
T_{\mathit{seek}} = \left(N_{\mathit{d}} \cdot \frac{d}{\min(r_{\mathit{tip}}, r_{\mathit{tilt}})}\right)
+ \left(N_{\mathit{h}} \cdot \frac{d}{r_{\mathit{tip}}}\right),
\end{equation}
where $r_\mathit{tip}$ and $r_\mathit{tilt}$ are distances in meters caused by the angular change of the FSM tip, $\omega_\mathit{tip}$, and tilt, $\omega_\mathit{tilt}$, rates.

The seeking optical terminal dwells at each step, $t_\mathit{dwell}$, to give the partner spacecraft a chance to start tracking. The process results in a total link acquisition delay defined as
\begin{equation}
T_{\mathit{acq}} = T_{\mathit{seek}} + N_{steps} \cdot t_{\mathit{dwell}}.
\end{equation}

\subsection{Closed-Loop Tracking Transition}

Once fine pointing is stabilized, the system transitions into closed-loop tracking mode, where each terminal maintains microradian-level alignment using continuous feedback from the track sensor.
FSMs correct for jitter, drift, thermal distortion, and platform vibrations by continuously monitoring the received signal strength and centroid offset from the tracking sensor, applying real-time adjustments to maintain alignment.

The control system does this by polling the signal strength of the track sensor at a frequency $f$ and incrementing a discrete-time variable if the signal strength exceeds a threshold, or decrementing the variable if the strength falls below the threshold.
The transition to closed-loop tracking is complete once the convergence threshold, $\alpha$, is reached.
The duration of the transition is then
\begin{equation}
    T_{acq-to-track} = \frac{N_{samples}}{f}.
\end{equation}

The value of $N_{samples}$ is estimated using the following probabilistic model that determines the number of samples needed to transition into closed-loop tracking:
\begin{equation}
    \alpha = (N_{samples} \cdot p_{signal}) - (2 \cdot N_{samples} \cdot (1-p_{signal})),
\end{equation}
where $p_\mathit{signal}$ represents the probability that the signal strength is above the threshold for a given timestep.

\section{PAT Delay Model Analysis}

To validate the proposed model, we generated simulations of thousands of contact opportunities using realistic orbits for LEO and deep space missions.
Open-source implementations of the optical link PAT delay model and the simulator are publicly available\footnote{https://github.com/jason-gerard/laser-link-scheduler}.
Table~\ref{tab:simulation_params} shows the selected parameters taken from a variety of real space missions.

\begin{table}[t]
\centering
\caption{Simulation parameters~\cite{israel_early_2023, boroson_overview_2014, riesing_operations_2023}.}
\label{tab:simulation_params}
\begin{tabular}{@{}lll@{}}
\toprule
Parameter & IPN & LEO \\ \midrule
Slew rate Az & 1.0 deg/s & 2 deg/s \\
Slew rate El & 1.0 deg/s & 0.5 deg/s \\
FSM tip & 5.0 mrad/s & 8.5 mrad/s \\
FSM tilt & 5.0 mrad/s & 8.5 mrad/s \\
Dwell time & 0.5 s & 0.5 s \\
Beam width & 0.2 deg & 0.2 deg \\
FOU & 1.0 deg & 0.75 deg \\
$f$ & 1 kHz & 1 kHz \\
$p_{signal}$ & 0.7 & 0.7 \\ \bottomrule
\end{tabular}
\end{table}

\begin{figure}[t]
    \centering
    \includegraphics[width=1.0\linewidth,height=\textheight,keepaspectratio]{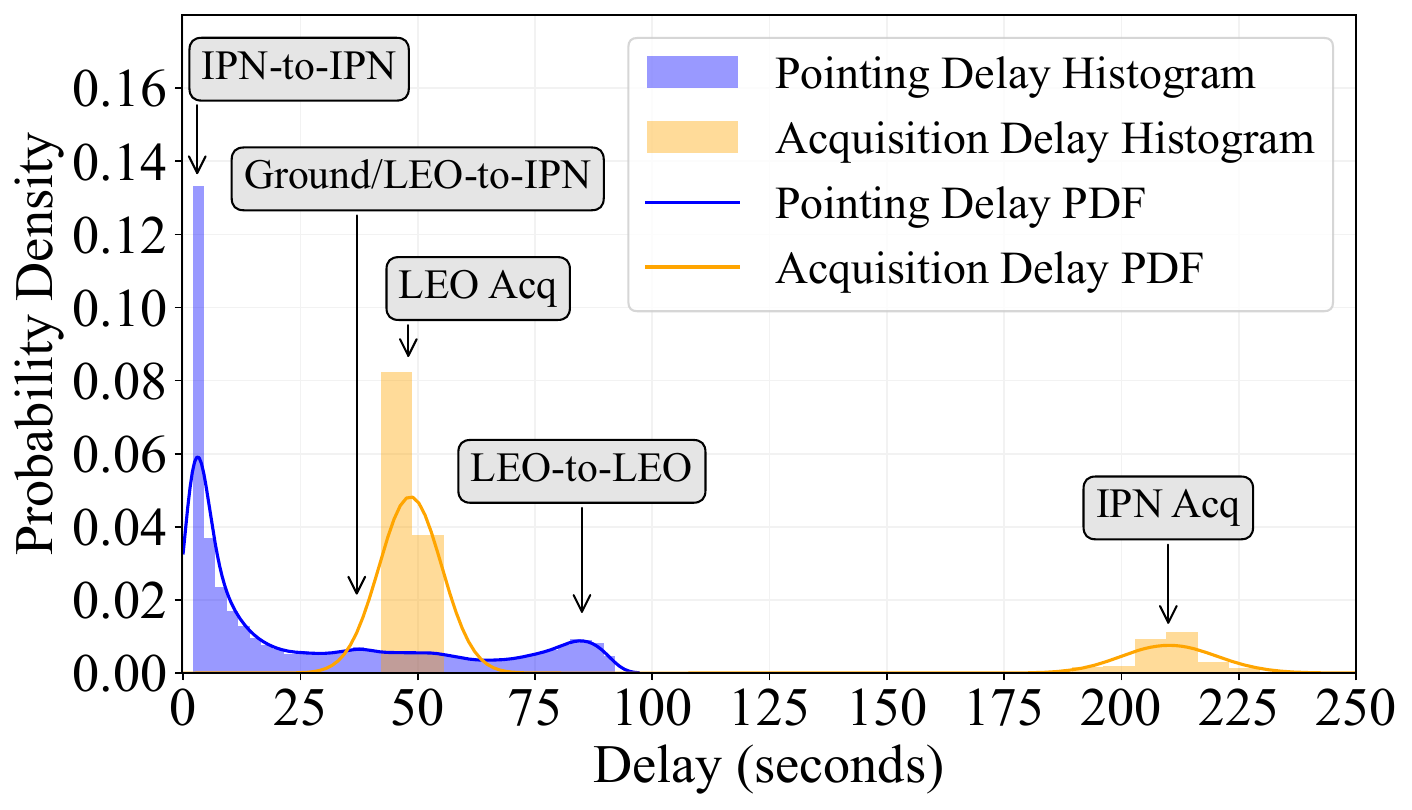}
    \caption{PDF overlayed with a histogram shows clear characterization of different types of optical links based on pointing and link acquisition delays.}
    \label{fig:retargeting_delay_pdf}
\end{figure}

Figure~\ref{fig:retargeting_delay_pdf} shows a probability density function (PDF) overlaid with histograms of pointing and link acquisition delays sampled from thousands of LEO and IPN contacts.
The distribution clearly illustrates the categorization of optical links based on the current and previous link characteristics, providing insight into how system geometry and prior state impact PAT delays. The bimodal nature of the acquisition delay distribution is evident, with LEO links (labeled as \textit{LEO Acq}) clustering around 50 seconds, whereas IPN links (labeled as \textit{IPN Acq}) exhibit a peak near 200 seconds. The latter reflects the longer beam search times driven by larger fields of uncertainty.

In contrast, the pointing delay distribution is trimodal.
The first sharp peak, labeled \textit{IPN-to-IPN}, at 3 seconds, corresponds to groups of current and previous IPN link pairs, where retargeting results in minimal slews due to aligned geometry.
The second peak, labeled \textit{Ground/LEO-to-IPN}, at 40 seconds, occurs when an Earth-orbiting relay satellite or optical ground station retargets toward a deep space optical terminal.
The third peak, labeled \textit{LEO-to-LEO}, at 80 seconds, corresponds to LEO satellites or ground stations slewing from LEO or GEO targets to other LEO satellites, requiring larger angular slews and additional stabilization time.

The structured multimodal behavior validates the underlying PAT delay model, demonstrating that prior link geometry and mission architecture significantly affect PAT delays and should be explicitly modeled in FSO network simulations to accurately predict contact planning and network capacity.
The figure highlights that while hardware capabilities set the lower bounds for PAT delays, operational context determines which regime the system operates in.
It is essential to consider constellation geometry, pointing constraints, and acquisition strategies in the design and evaluation of scalable optical networking architectures.

Figure~\ref{fig:retargeting_delay_pdf} shows that IPN links are characterized by short slews, due to the relative proximity of deep space assets within the FOR, and long acquisition delays, driven by the larger FOU.
IPN acquisition accounts for around 90\% of the total duration of PAT, depending on whether the previous link was an IPN link or a LEO link.
LEO links are categorized by three primary types: along-track (static, low retargeting frequency), cross-track (frequently retargeted), or LEO-to-ground (also frequently retargeted).
Overall, LEO links have shorter acquisition delays and long slews when transitioning from LEO to IPN links, which can vary between 90 and 180 degrees of total rotation.
Along-track and cross-track links both have around double the PAT delay of LEO-to-ground.

The modeled LEO acquisition delays of 30 to 60 seconds align with NASA TBIRD mission results, which reported similar acquisition times under comparable FOU and tip/tilt rates~\cite{riesing_operations_2023}.
On average, link acquisition for TBIRD took up to 20\% of the 5-minute contact window, with data rates tapering off at the start and end.
For IPN cases, the modeled 200-second acquisition delays match NASA's DSOC optical terminal tests, which observed acquisition times between 180 and 240 seconds~\cite{biswas_deep_2024}.
These comparisons demonstrate that the PAT delay model accurately reflects real mission performance across both LEO and deep space regimes.

\begin{figure}[t]
    \centering
    \includegraphics[width=0.7\linewidth,height=\textheight,keepaspectratio]{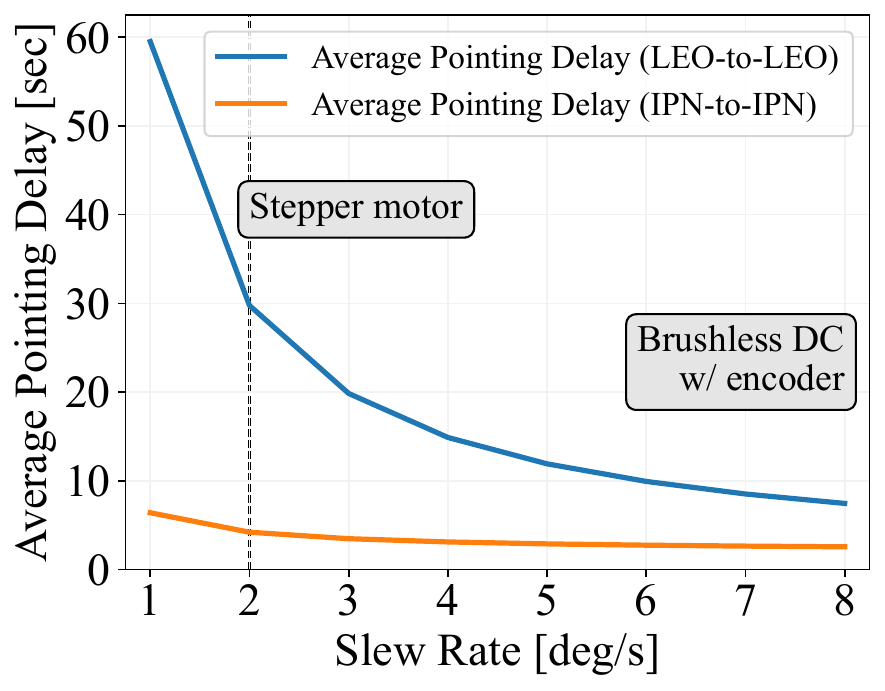}
    \caption{Average pointing delay increases exponentially for LEO-to-LEO links as CPA slew rate decreases, while IPN links have little difference.}
    \label{fig:avg_pointing_delay}
\end{figure}

Figure~\ref{fig:avg_pointing_delay} shows that the average pointing delay for LEO-to-LEO links decreases with increasing slew rate, though the relationship is nonlinear, with diminishing returns at higher rates.
Moving from lower slew rates (e.g., 1 to 2~deg/s with stepper motors) to moderate rates (4~deg/s) reduces pointing delays from 50 to 15 seconds, demonstrating the benefit of faster actuators for retargeting-heavy networks.
Above 4~deg/s, using brushless DC motors can achieve pointing delays near 6 seconds, but further gains diminish as settling time and control loop stability begin to dominate.
The higher SWaP of these actuators may not be justified for missions with infrequent retargeting, but can be worthwhile for LEO mega-constellations where frequent re-targeting benefits from reduced PAT delays, improving network utilization.
In contrast, IPN-to-IPN links show minimal delay reduction with increasing slew rate due to smaller slew angles from the close relative proximity of deep space assets.

\begin{figure}[t]
    \centering
    \includegraphics[width=0.8\linewidth,height=\textheight,keepaspectratio]{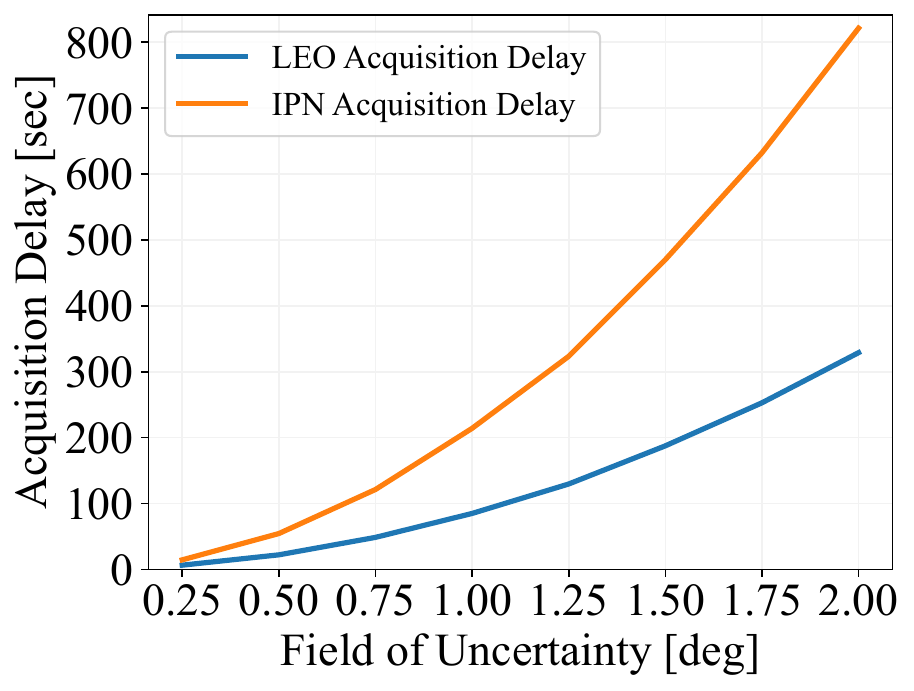}
    \caption{Both LEO and IPN acquisition delays increase exponentially as the FOU increases.}
    \label{fig:acq_delay_by_fou}
\end{figure}

Figure~\ref{fig:acq_delay_by_fou} demonstrates how acquisition delay increases sharply with the FOU, highlighting the exponential sensitivity of acquisition time to initial pointing uncertainty in FSO networks.
As the FOU expands from 0.25 to 2 degrees, the time required to complete beam searching and acquire the link grows rapidly, with delays for interplanetary (IPN) links exceeding 800 seconds at the upper bound. 
Meanwhile, LEO acquisition, though faster overall, still exhibits significant growth in delay with increasing FOU.
This behavior aligns with analytical models where acquisition delay scales with the search area divided by the beam area, leading to an effectively exponential relationship under practical system parameters.

These results underscore the criticality of minimizing initial pointing uncertainty through high-precision attitude knowledge and prediction, as even modest improvements in FOU can yield substantial reductions in acquisition time, improving contact efficiency and network throughput.
While pure star tracker-based approaches or cooperative ephemeris-sharing can theoretically eliminate the need for extended beam searching and reduce link acquisition delays, such methods introduce complexity and cost, particularly when scaling to deep space constellations where maintaining precise absolute position knowledge for every node may be infeasible.
This analysis suggests that while LEO optical networks inherently benefit from shorter acquisition times, all types of FSO networking can benefit from investments in reducing FOU, trading initial system cost and complexity for operational gains in acquisition efficiency and increased link uptime.

\section{Conclusion}
Free-space lasers enable deep space missions to return more scientific data by providing high-capacity network links.
However, the primary drawback of optical communications is the delay required for pointing and acquisition to establish each link.
Existing work often overlooks these PAT delays, making it difficult to apply theoretical scheduling and routing algorithms in practice.
This paper presented delay models for optical link pointing and acquisition with tunable parameters, allowing characterization of different link types to evaluate algorithm performance in optical networks.
We validated the model against real-world mission data, demonstrating that PAT delays can account for a significant portion of contact time, particularly when transitioning from LEO to interplanetary links.
In future work, we will apply this model to optical link scheduling algorithms to further understand and mitigate the impact of PAT delays in optical satellite networks.

\section*{Acknowledgment}
The authors would like to acknowledge the support from Concordia's PhD Fellowship Award, NSERC Canada Discovery Grant RGPIN-2024-05730, ANID Basal Project AFB240002, and the French National Research Agency (ANR) under project ANR-22-CE25-0014-01.

\bibliographystyle{IEEEtran}
\bibliography{references}

\end{document}